\documentclass{article}
\usepackage{spconf,amsmath,graphicx}
\usepackage{enumitem}
\usepackage[page]{appendix}
\PassOptionsToPackage{hyphens}{url}\usepackage{hyperref}
\usepackage{listings}
\hypersetup{colorlinks=true}

\usepackage{amsfonts}
\usepackage{booktabs,multirow}
\usepackage{adjustbox}
\usepackage{algorithm}
\usepackage{algpseudocode}
\usepackage{color}
\usepackage{xcolor}

\newcommand{\vct}[1]{\boldsymbol{\mathbf{#1}}} 
\newcommand{\T}{^{\mathsf{T}}} 
\newcommand{\mat}[1]{\boldsymbol{\mathbf{#1}}} 
\definecolor {CodeBackground}   {rgb}{0.95,  0.95,  0.95}

\title{Turn-to-Diarize: Online speaker diarization constrained by transformer transducer speaker turn detection}
\name{Wei Xia$^*$, Han Lu$^*$, Quan Wang$^*$, Anshuman Tripathi, Yiling Huang, Ignacio Lopez Moreno, Hasim Sak\thanks{* Equal contribution. Wei performed this work as an intern at Google.}}
\address{Google LLC, USA
\\[4pt] {
  \normalsize
   \{
    \href{mailto:ericwxia@google.com}{\nolinkurl{ericwxia}},
    \href{mailto:luha@google.com}{\nolinkurl{luha}},
    \href{mailto:quanw@google.com}{\nolinkurl{quanw}},
    \href{mailto:anshumant@google.com}{\nolinkurl{anshumant}},
    \href{mailto:yilinghuang@google.com}{\nolinkurl{yilinghuang}},
    \href{mailto:elnota@google.com}{\nolinkurl{elnota}},
    \href{mailto:hasim@google.com}{\nolinkurl{hasim}}
    \}
  {\tt @google.com}
}}

\begin{document}
\ninept
\maketitle
\begin{abstract}
In this paper, we present a novel speaker diarization system for streaming on-device applications. In this system, we use a transformer transducer to detect the speaker turns, represent each speaker turn by a speaker embedding, then cluster these embeddings with constraints from the detected speaker turns.
Compared with conventional clustering-based diarization systems, our system largely reduces the computational cost of clustering due to the sparsity of speaker turns. Unlike other supervised speaker diarization systems which require annotations of time-stamped speaker labels for training, our system only requires including speaker turn tokens during the transcribing process, which largely reduces the human efforts involved in data collection.
\end{abstract}
\section{Introduction}
\label{sec:intro}

Speaker segmentation is a key component in most modern speaker diarization systems~\cite{park2021review}. The outputs of speaker segmentation are usually short segments which can be assumed to consist of individual speakers. With these homogeneous segments, we can extract speaker embedding such as i-vector~\cite{dehak2011front} or d-vector/x-vector~\cite{ge2e,snyder2018x} from each segment to represent its speaker identify. The speaker embeddings can be either directly clustered with conventional clustering algorithms such as K-means~\cite{shum2013unsupervised} or spectral clustering~\cite{wang2017speaker}, or fed into a supervised model such as unbounded interleaved-state recurrent neural networks (UIS-RNN)~\cite{zhang2019fully}, discriminative neural clustering (DNC)~\cite{li2019discriminative}, or permutation-invariant training~\cite{fujita2019end,e2ediarizationpatent}.

There are typically three approaches to the speaker segmentation problem: 
\begin{enumerate}[leftmargin=*]
    \item Uniform speaker segmentation: The entire utterance is divided into segments of uniform length. Although this approach is simple and easy to implement, it is difficult to find a good segment length --- long segments may very likely contain speaker turn boundaries, while short segments carry insufficient speaker information. For example, the systems described in~\cite{wang2017speaker,zhang2019fully} are based on segments of a fixed length of 400ms, while the system described in~\cite{garcia2017speaker} is based on segments of 2s.
    \item ASR-based word segmentation: Automatic speech recognition (ASR) models generate word boundaries, which could be used as word-level segmentation. Although we can usually safely assume that a word-level segment comes from a single speaker, word segments are still too short to carry sufficient speaker information.
    \item Supervised speaker turn detection (\emph{a.k.a.} speaker change detection): A dedicated model is trained to detect the exact timestamps of speaker turns, such as the systems described in~\cite{yin2017speaker,yin2018neural}.
\end{enumerate}

Apparently, among the above three approaches, supervised speaker turn detection has multiple advantages. First, since a segment covers a full continuous speaker turn, it carries sufficient information to extract robust speaker embeddings. Besides, for very long conversational speech, the number of speaker turns is usually much smaller than the number of appropriate fixed-length short segments --- this would largely reduce the computational cost of clustering the segment-wise embeddings.

The speaker turn detection models described in~\cite{yin2017speaker,yin2018neural} are purely based on acoustic information of the training utterances. These kinds of models fail to leverage the rich semantic information in the data. For example, by only looking at the text transcript of the conversation ``How are you I'm good'', we can confidently conjecture there is a speaker change between ``How are you'' and ``I'm good''. In~\cite{park2018multimodal}, both lexical and acoustic cues are considered to estimate the speaker turn for diarzation with ASR outputs. 

Although many recent end-to-end speaker diarization systems have shown very promising results~\cite{fujita2019end,maiti2021end,e2ediarizationpatent}, these systems usually require a large amount of carefully annotated training data. The annotation process usually requires the human annotator to assign accurate \emph{timestamps} to the speaker turns, and manually identify different speakers across these turns. Our internal study shows that this kind of annotation process takes roughly 2 hours for a single annotator to annotate 10 minutes of audio for one pass.

In this paper, we propose Transformer Transducer~\cite{zhang2020transformer,tripathi2020transformer,yeh2019transformer} based speaker turn detection that is jointly trained with the ASR model~\cite{he2019streaming}. We use a special token \texttt{<st>} to represent the speaker turn, and inject this token into ground truth transcripts of the ASR training data. This approach not only makes better use of semantic information in the speech data, but also reduces annotation costs --- annotating the \texttt{<st>} token as part of transcribing the speech data is much easier than annotating the exact timestamp of the speaker turn events.

Our work shares similarities with the system proposed in~\cite{shafey2019joint}. In~\cite{shafey2019joint}, two role-specific tokens \texttt{<spk:dr>} and \texttt{<spk:pt>} are injected to the ASR transcripts to indicate speech from the doctor and the patient. However, such a system cannot be used for generic speaker diarization problems where: (1) There could be more than 2 speakers; (2) The speakers are not constrained to specific roles.

The original contributions of this paper include: (1) We proposed an efficient speaker diarization system for streaming on-device applications that do not rely on expensive timestamped annotations; (2) We proposed a transformer transducer-based model for joint ASR and speaker turn detection; (3) We proposed a constrained spectral clustering algorithm that incorporates the prior information from speaker turns into the spectral clustering process.

\section{Methods}

\subsection{System architecture}

\begin{figure}
	\centering
	\includegraphics[width=\linewidth]{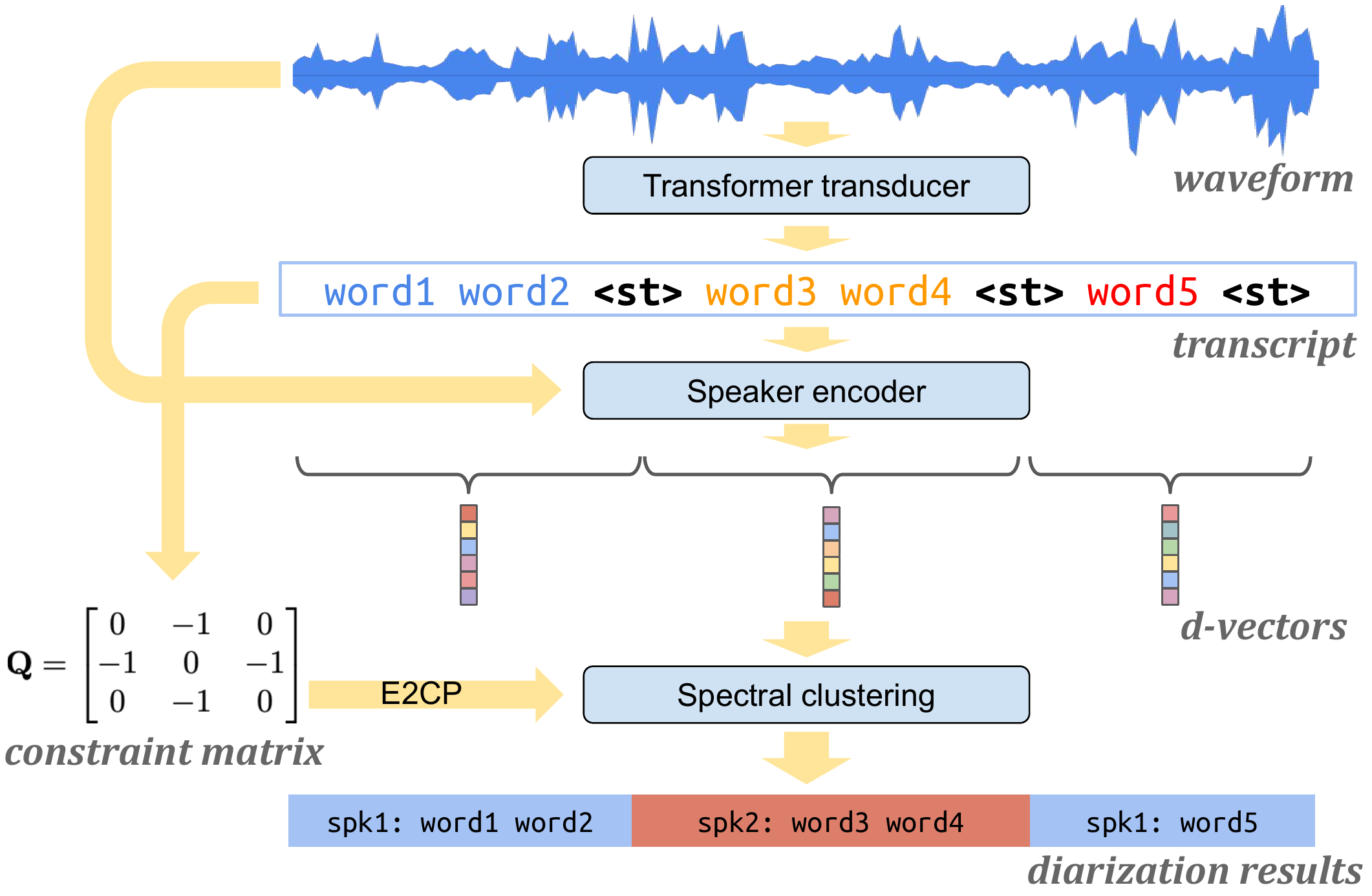}
	\caption{System architecture of the speaker diarization system.}
	\label{fig:inference}
\end{figure}

The system architecture is shown in Fig.~\ref{fig:inference}. The input utterance is first fed into a transformer transducer model for joint ASR and speaker turn detection.
Then the utterance is segmented into speaker turns, and each turn is fed into an LSTM based speaker encoder to extract a d-vector embedding.
We use a spectral clustering algorithm to cluster these turn-wise d-vectors, but with constraints from the detected speaker turns\footnote{We open sourced the constrained spectral clustering algorithm at \url{https://github.com/wq2012/SpectralCluster}}.

\begin{table}[t]
    \centering
    \caption{Hyper-parameters of a Transformer block.}
    \begin{tabular}{c|c}
    \toprule
    Input feature projection &  256 \\
    Dense layer 1 & 1024 \\
    Dense layer 2 & 256 \\
    Number attention heads & 8 \\
    Head dimension & 64  \\ 
    Dropout ratio & 0.1  \\
    \bottomrule
    \end{tabular}
    \label{table:scd-arch}
\end{table}

\subsection{Online diarization}

Since both the transformer transducer-based speaker turn detection model and the LSTM based speaker encoder model are streaming models, the bottleneck of latency is the clustering algorithm. Previous studies have shown that online clustering algorithms such as Links~\cite{mansfield2018links} are significantly worse than offline clustering algorithms such as spectral clustering~\cite{wang2017speaker}. To have both great performance and low latency at the same time, we use spectral clustering in an online fashion: every time when we have a new speaker embedding, we run spectral clustering on the entire sequence of all existing embeddings. Because speaker embeddings are extracted from speaker turns which are usually sparse, the sequence is usually relatively short even for hour-long conversations, thus making the clustering inexpensive to run and feasible for on-device deployment.

\subsection{Transformer transducer based speaker turn detection}

\label{sec:t-t}

Recurrent neural network transducer (RNN-T)~\cite{graves2012sequence} is an ASR model architecture that can be trained end-to-end with RNN-T loss. Such architecture includes an audio encoder, a label encoder, and a joint network that produces the final output distribution over all possible labels. We adopt Transformer Transducer (T-T) ~\cite{zhang2020transformer}, a variant of the RNN-T model, as the speaker turn detection model for its advantages of faster inference speed, and no long-form deletion issues. We also use a bigram label encoder proposed in~\cite{tripathi2020transformer} to further speed up the decoding via an embedding table lookup.

To create training targets, inspired by~\cite{shafey2019joint} that adds speaker roles as part of the transcript (\emph{e.g.} ``hello how are you \texttt{<spk:dr>} I am good \texttt{<spk:pt>}''), we add a special speaker turn token \texttt{<st>} between two different speakers' transcripts (\emph{e.g.} ``hello how are you \texttt{<st>} I am good \texttt{<st>}'') to model speaker turns during training. Compared to audio-only models~\cite{yin2017speaker,yin2018neural}, this model can potentially utilize the language semantics as a signal for speaker segmentation. T-T is trained in a sequence-to-sequence fashion, where the source is log-Mel filterbank energy features, and the target is the transcript that includes both transcript texts and the special speaker turn tokens. At inference time, we ignore all the texts output by the model except for the \texttt{<st>} tokens, and their corresponding timestamps. These timestamps are later used as the speaker boundaries in the diarization system.

We train the T-T model using Fisher~\cite{cieri2004fisher}, the training subset of Callhome American English~\cite{canavan1997callhome}, and an internal dataset collected from around 7500 hours of YouTube videos. Training utterances are segmented into 15 seconds segments with speaker turn tokens added so it fits into the memory more easily. The audio encoder has 15 layers of Transformer blocks. Each block has 32 left context and no right context. The hyper-parameters for each repeated block can be found in Table~\ref{table:scd-arch}. We also use a stacking layer after the second transformer block to change the frame rate from 30ms to 90ms,  and an unstacking layer after the 13th transformer block to change the frame rate from 90ms back to 30ms, to speed up the audio encoder as proposed in~\cite{tripathi2020transformer}. The bigram label encoder embedding table has embeddings with a dimension of 256. For the joint network, we have a projection layer that projects the audio encoder output to 256-d. At the output of the joint network, it produces a distribution over 75 possible graphemes with a softmax layer. For optimization, we follow the same hyper-parameters described in~\cite{zhang2020transformer}.

\subsection{Speaker encoder}

\label{sec:sid}

Our speaker encoder is a text-independent speaker recognition model
trained with the generalized end-to-end extended-set softmax loss~\cite{ge2e,pelecanos2021dr}. The speaker encoder model has 3 LSTM layers each with 768 nodes and a projection size of 256. The output of the last LSTM layer is then linearly transformed to the final 256-dimension d-vector. The same model was used in~\cite{rikhye2021personalized}.

At inference time, we use the detected speaker turns as signals to reset the LSTM states of the speaker encoder, such that it does not carry information across different turns. For each speaker turn, we use the embedding at roughly 75\% of this turn to represent this speaker turn, such that it has sufficient information from this turn, and is not too close to the speaker boundary which could be inaccurate or contain overlapped speech. Besides, because speaker turn detection may have false rejections, to reduce the risk, we further segment any turns that are longer than 6 seconds. This type of segmentation is also used to construct ``Must-Link'' constraints as described in Sec.~\ref{sec:constraint}.

\subsection{Spectral clustering}
\subsubsection{Recap of spectral clustering}
We use the spectral clustering method~\cite{von2007tutorial} to predict speaker labels on turn-wise speaker embeddings. Given a set of $N$ data samples $\{\vct{x}_1, \vct{x}_2, ..., \vct{x}_N\}$, we can construct a similarity graph by computing pairwise similarities $a_{ij}$. Let $\mat{A}\in \mathbb{R}^{N\times N}$ be the affinity matrix of the graph, the affinity of two samples $\vct{x}_i$ and $\vct{x}_j$ is $a_{ij} = \frac{1}{2}(1 + \cos(\vct{x}_i, \vct{x}_j)) $. 
Since spectral clustering is sensitive to the quality and noise of a similarity graph, we define the following refinement operations~\cite{wang2017speaker} on the affinity matrix to model the local neighborhood relationships between data samples. 
\begin{enumerate}[leftmargin=*]
    \item Row-wise soft-thresholding with $p$-percentile: affinity values that are larger than the $p$-percentile of the row are binarized to 1; affinity values that are smaller than the $p$-percentile are multiplied with $0.01$.
    \item We apply an average symmetrization operation to make the affinity matrix symmetric: 
        $\hat{\mat{A}} = \frac{1}{2}(\mat{A} + \mat{A}\T)$
\end{enumerate}
Gaussian Blur operation~\cite{wang2017speaker} is also applied at the beginning for window-wise dense embeddings to smooth and denoise the data.
We find that the diarization performance is significantly affected by the hyper-parameter $p$ for the $p$-percentile. The default $p$ is 0.95 in our experiments. As proposed in~\cite{park2019auto}, we can use a ratio value $r(p)$ as a good proxy of the Diarization Error Rate (DER). Here, we define the ratio $r(p) = \frac{\sqrt{1-p}}{g_p}$, where $g_p=\underset{k}{\max} \frac{\lambda_{p,k+1}}{\lambda_{p,k} + \epsilon}$ is the maximum eigen-gap given $N$ eigen-values $\{\lambda_{p,1}, \lambda_{p,2},..., \lambda_{p,N}\}$ of the refined affinity matrix, and $\epsilon$ is a very small value.
By minimizing this ratio $r(p)$, we can automatically select an appropriate $p$-percentile for our spectral clustering algorithm without a holdout development set. 
Note that for every search step, the auto-tuning method requires an eigen-decomposition operation of the affinity matrix,  which is the major bottleneck of the clustering algorithm. If the speaker embeddings are dense, the size of the affinity matrix and therefore the computational cost become very large.

Next we define a graph Laplacian matrix $\mat{L}$. Given an affinity matrix $\mat{A}$, the degree matrix $\mat{D}$ is a diagonal matrix where the diagonal elements $d_{ii}=\sum_{j=1}^{N}a_{ij}$. The unnormalized Laplacian matrix $\mat{L} = \mat{D} - \mat{A}$, and the normalized Laplacian $\bar{\mat{L}}  = \mat{D}^{-1/2}\mat{L}\mat{D}^{-1/2}$. In our experiments, we use the normalized graph Laplacian $\bar{\mat{L}}$. To perform spectral clustering,
\begin{itemize}[leftmargin=*]
    \item We apply eigen-decomposition and estimate the speaker number $k$ using the maximum eigengap method.
    \item We choose the first $k$ eigen-vectors and apply a row-wise re-normalization of the spectral embeddings. K-means algorithm is applied on the spectral embeddings to predict class labels.
\end{itemize}

\subsubsection{Speaker turn priors}
Given a sequence of speaker segments and speaker turn information, we know that two \textit{neighboring} segments adjacent to the \texttt{<st>} token are from different speakers. 
Therefore, speaker turn prior information can be used as pairwise constraints to guide the clustering process. Pairwise constraints, unlike the class labels of data, do not provide explicit class information and are considered a weaker form of supervisory information. Using the proposed T-T speaker turn detection in Sec.~\ref{sec:t-t}, we can predict a confidence score for each speaker turn token \texttt{<st>}. The general objective is to encourage speaker labels of segments across high confidence \texttt{<st>} to be different and speaker labels of segments without \texttt{<st>} token or across very low confidence \texttt{<st>} to be the same.

\subsubsection{Spectral clustering with pairwise speaker turn constraints}
\label{sec:constraint}
With the pairwise constraints from speaker turn side information, we can perform a constrained spectral clustering that tries to find a partition (or multiple partitions) that maximizes constraint satisfaction and minimizes the cost on the similarity graph.

Let $\mat{Q}\in \mathbb{R}^{N\times N}$ be a constraint matrix. If there is a speaker turn between segment $i$ and $i+1$, and the confidence of the \texttt{<st>} token $c(\texttt{<st>})$ is larger than a threshold $\sigma$, we define this pair as a ``Cannot-link'' (CL)~\cite{basu2008constrained}. If there is no speaker turn between two segments, we define it as a ``Must-Link'' (ML). $\mat{Q}_{ij} = 0$ if $i, j$ are not neighboring segments.
\begin{equation}
\label{eq:initial_constraint}
\mat{Q}_{ij}=
\begin{cases}
{-1, }& \text{If}\; (i, j) \in \text{CL} \; \text{and}\;  c(\texttt{<st>}) > \sigma ; \\
{+1, }& \text{If}\, (i, j) \in \text{ML} ; \\
{0, }& {\text{Otherwise}.}
\end{cases}    
\end{equation}
The generated constraint matrix is a banded sparse matrix since we only have limited speaker turns. To fully utilize the inherent information from the speaker turns, we can infer more constraint information using the Exhaustive and Efficient Constraint Propagation (E2CP) method in~\cite{lu2013exhaustive}.

First, we divide the pairwise constraint propagation problem into a two-class label propagation sub-problem, where we treat an ML pair as a positive class and a CL pair as a negative class. 
The class labels are propagated in vertical and horizontal directions respectively. Let $\bar{\mat{A}} = \mat{D}^{-1/2} \mat{A} \mat{D}^{-1/2}$, which is the symmetric regularization of the unrefined affinity $\mat{A}$. 
$\mat{Z}$ is the initial constraint matrix defined in Eq.~\ref{eq:initial_constraint}.
A parameter $\alpha$ is used to control the relative amount of constraint information from its neighbors and the initial constraints. It is set to 0.4 in our experiments. We perform vertical propagation first until the convergence and then the horizontal propagation. By combining these two propagations, we diffuse the pairwise constraints to the whole graph. 
With the E2CP algorithm, the final propagated constraint matrix $\mat{Q}^{*}$ has a closed-form feasible solution, which is formulated as below,
\begin{align}
\mat{Q}^{*} = (1-\alpha)^{2} (I-\alpha \bar{\mat{A}})^{-1} \mat{Z}(I-\alpha \bar{\mat{A}})^{-1} .
\end{align}
Using this propagated constraint matrix $\mat{Q}^{*}$, we can obtain an adjusted affinity matrix $\hat{\mat{A}}$, where
\begin{align}
\hat{\mat{A}}_{ij}= \begin{cases}
{1-\left(1-\mat{Q}_{ij}^{*}\right)\left(1 - \mat{A}_{ij}\right), }& {\text{If}\; \mat{Q}_{ij}^{*} \geq 0 \, ;}\\
 {\left(1+\mat{Q}_{ij}^{*}\right) \mat{A}_{ij}, }& {\text{If}\; \mat{Q}_{ij}^{*}<0 \, .}\\
\end{cases}
\end{align}
For constraint $\mat{Q}_{ij} > 0$, it increases the similarity between the sample $\vct{x}_i$ and $\vct{x}_j$; if the constraint is negative, the similarity is decreased. After this operation, we still perform the normalized Laplacian matrix based spectral clustering to predict cluster labels\footnote{To summarize, with E2CP, the workflow is: affinity $\rightarrow$ constraint $\rightarrow$ refinement $\rightarrow$ Laplacian.}.

\begin{table*}[t]
  \centering
  \caption{Confusion (\%), total DER (\%) and GFLOPS/s on three datasets for different embeddings and methods.}
    \begin{adjustbox}{max width=1\linewidth}
    \begin{tabular}{l|l|cc|cc|cc|c|c}
    \toprule
    \multicolumn{1}{c|}{\multirow{2}{*}{System}} & \multicolumn{1}{c|}{\multirow{2}{*}{Method}} & 
    \multicolumn{2}{c|}{Inbound} & \multicolumn{2}{c|}{Outbound} & \multicolumn{2}{c|}{Callhome Eval} &
    \multicolumn{1}{c|}{GFLOP/s} &
    \multicolumn{1}{c}{GFLOP/s} \\
    \cline{3-8} & \multicolumn{1}{c|}{} & 
    \multicolumn{1}{c}{Conf.} & \multicolumn{1}{c|}{DER} & \multicolumn{1}{c}{Conf.} & \multicolumn{1}{c|}{DER} & \multicolumn{1}{c}{Conf.} & \multicolumn{1}{c|}{DER}  &
    at 10min & at 1h\\
    \hline
    \multicolumn{1}{c|}{\multirow{2}{*}{Dense d-vector}} 
        & Dense                   & 17.98 & 22.13 & 10.66 & 15.97 & 5.39  & 7.76 & 0.85 & 36.54 \\
        & Dense + Auto-tune       & 14.09 & 18.24 & 9.56  & 14.88 & 5.42  & 7.79 & 4.76 & 361.37 \\
    \hline
    \multicolumn{1}{c|}{\multirow{4}{*}{Turn-to-diarize}} 
        & Turn                    & 17.87 & 19.43 & 8.41  & 10.34 & 8.23  & 10.08  & 1.00 & 1.18 \\
        & Turn + E2CP             & 17.21 & 18.77 & 7.94  & 9.86  & 3.56  & 5.41  & 1.00 & 1.18 \\
        & Turn + Auto-tune        & 13.83 & 15.39 & 7.01  & 8.93  & 5.11  & 6.95  & 1.02 & 2.81 \\
        & Turn + E2CP + Auto-tune & \textbf{13.66} & \textbf{15.22} & \textbf{6.86} & \textbf{8.78} & \textbf{3.49}  & \textbf{5.33} & 1.02 & 2.81 \\
    \bottomrule
    \end{tabular}
    \end{adjustbox}
  \label{tab:DER_results}
\end{table*}

\section{Experiments}

\subsection{Data and metrics}

The training data for the speaker turn detection model and the speaker encoder model have been described in Sec.~\ref{sec:t-t} and Sec.~\ref{sec:sid}, respectively. To evaluate our speaker diarization system, we use a vendor-provided call center domain dataset. This dataset consists of anonymized utterances containing telephone conversations between call center attendants and customers, and can be divided into two subsets:
\begin{enumerate}[leftmargin=*]
    \item The ``Outbound'' subset, which includes 450 conversations initiated by the call center. This dataset has approximately 35 hours of speech in total. Each utterance has 2 speakers.
    \item The ``Inbound'' subset, which includes 250 conversations initiated by customers. This dataset has approximately 22 hours of speech in total. Each utterance has 2 to 10 speakers.
\end{enumerate}

Apart from the internal call center domain dataset, we also evaluate our diarization system on the Callhome American English data (LDC97S42)~\cite{canavan1997callhome}. The Callhome American English corpus is divided into the train, dev, and eval sets. As the train subset has been used for training the speaker turn detection model, we report the diarization results on the eval set of 20 utterances, which is about 1.7 hours of recordings in total.

We report the confusion errors and total Diarization Error Rate (DER) computed with the pyannote.metrics library~\cite{bredin2017pyannote}, and follow the same evaluation protocols as~\cite{wang2017speaker,zhang2019fully}. 

\subsection{Experimental results}







We use the speaker diarization system described in~\cite{wang2017speaker} as our baseline system (using the same speaker encoder model as described in Sec.~\ref{sec:sid}), and refer to it as the ``dense d-vector'' system, as the speaker embeddings are extracted from 400ms short segments.
In Table~\ref{tab:DER_results}, we show the experimental results of the ``dense d-vector'' and the proposed ``turn-to-diarize'' systems on the Internal Inbound, Outbound datasets, as well as the publicly available Callhome evaluation set. 
We report the total Diarization Error Rate (DER) and speaker Confusion Error Rate. The remaining errors are from False Alarm (FA) and Miss which are mostly caused by the Voice Activity Detection. 
As shown in the table, the turn-to-diarize method achieves better diarization results on the Inbound, and Outbound datasets, compared with the dense d-vector system. There is a relative 12.20\% and 35.25\% reduction in DER on the Inbound and Outbound datasets respectively. It shows that longer-duration speaker turn embeddings that capture more speaker characteristics might be more useful for diarization.

Moreover, the spectral clustering algorithm relies on the quality of the similarity graph. We find that pruning small and noisy values with the hyper-parameter $p$-percentile is essential to construct a good graph. It significantly impacts the diarization performance. The auto-tuning method~\cite{park2019auto} based on the ratio $r(p)$ does not require a holdout development set to tune the hyper-parameters, and we use this method to automatically select a good $p$-percentile. For all three datasets, the $p$-percentile search range is from 0.4 to 0.95 with a step size of 0.05. 
The auto-tuning method is tuned per-utterance and requires one operation of eigen-decomposition at each search step. When we use the dense d-vector method, however, the size of the Laplacian matrix is very high, so the computational cost of the eigen-decomposition operation is much more expensive and it causes much larger latency compared with the turn-to-diarize method.

Another advantage of the turn-to-diarize method is that we can use speaker turn prior information as pairwise constraints to guide the clustering process. For the dense d-vector approach, a single segment may cross the speaker turn boundary. It may contain speech from two speakers and therefore causes more confusion errors.
Comparing the ``Turn'' only and ``Turn + E2CP'' methods, we can observe a relative 3.40\%, 4.64\%, and 46.33\% reduction of DER on the Inbound, Outbound, and Callhome datasets respectively. It indicates the detected speaker turns are not only useful for segmenting the input, but also helpful to constrain the spectral clustering. Moreover, we notice that the E2CP method usually works better when a good $p$-percentile is not selected. If the similarity graph is already well-constructed with a good $p$-percentile, the effect of the E2CP method is marginal.

The auto-tuning method can also be combined with the speaker turn constraints. In Table~\ref{tab:DER_results}, the combination of E2CP and auto-tune with the turn-to-diarize system achieves the best results on all three datasets. It consistently improves the performance of the ``Turn'' only method and the best results of the dense system by a large margin, indicating the effectiveness of our proposed methods.

\subsection{Computational cost}
\label{sec:comp_cost}
In Table~\ref{tab:DER_results} we also include floating-point operations per second (FLOP/s) analysis for each system after running for 10min and 1h. This analysis assumes: dense d-vector is based on 400ms segments; average speaker turn length is 4s; the average number of speakers is 4; auto-tune searches for 10 values of $p$; and clustering runs every 4s. As we can see, turn-to-diarize is dominated by the speaker turn detection (578 MFLOP/s) and speaker encoder (415 MFLOP/s) neural networks, and the costs of eigen-decomposition, E2CP, Laplacian and K-Means are almost negligible even after processing 10min of audio. For dense d-vector, the computational cost (mostly eigen-decomposition) significantly increases when the sequence grows and becomes unacceptable.

\section{Conclusions}

We proposed a speaker diarization system for streaming on-device applications.
The system is based on a transformer transducer model for joint speech recognition and speaker turn detection. The detected speaker turns are not only used to segment the input, but also to constrain the spectral clustering of the turn-wise speaker embeddings. By clustering turn-wise embeddings instead of short segment-wise embeddings, we significantly reduced computational cost, and achieved offline performance with online latency. One future work is to retrain our transformer transducer on multilingual datasets to make our speaker diarization system language independent, and to use visual signals as additional constraints when they are available.

\newpage
\bibliographystyle{IEEEbib}
\bibliography{refs}

\newpage
\begin{appendices}

\section{Feature frontend}
We used a shared feature frontend for the speaker turn detection model in Section~\ref{sec:t-t} and the speaker encoder model in Section~\ref{sec:sid}. This frontend first applies automatic gain control~\cite{prabhavalkar2015automatic} to the input audio, then extracts 32ms-long Hanning-windowed frames with a step of 10ms. For each frame, 128-dimensional log Mel-filterbank energies (LFBE) are computed in the range between 125Hz and 7500Hz. These filterbank energies are then stacked by 4 frames and subsampled by 3 frames, resulting in final features of 512 dimensions with a frame rate of 30ms. These features are then filtered by a CLDNN based Voice Activity Detection (VAD)~\cite{zazo2016feature} before fed into the speaker turn detection and the speaker encoder models.

\section{Additional details on speaker encoder model}

We introduced our LSTM-based speaker encoder model in Section~\ref{sec:sid}.
The training data of this model include a vendor collected multi-language speech query dataset covering 37 locales, as well as LibriVox,
CN-Celeb~\cite{fan2020cn},
TIMIT~\cite{garofolo1993darpa},
and VCTK~\cite{yamagishi2019cstr}.
Multi-style training (MTR)~\cite{lippmann1987multi,ko2017study,kim2017generation} with SNR ranging from 3dB to 15dB is applied during the training process for noise robustness. The same speaker encoder model was also used in~\cite{rikhye2021personalized,rikhye2021multi}.

\section{Exhaustive and Efficient Constraint Propagation approach}
The Exhaustive and Efficient Constraint Propagation (E2CP)~\cite{lu2013exhaustive} method is described in Algorithm~\ref{alg:e2cp}.

\begin{algorithm}
\caption{Exhaustive and Efficient Constraint Propagation (E2CP) method}\label{alg:e2cp}
\begin{algorithmic}
\Require Initial constraint matrix $\mat{Z}=\mat{Q}(0)$,  matrix $\bar{\mat{A}}$, propagation parameter $\alpha$.
\While{$\mat{Q}_{v}(t)$ not converged to $\mat{Q}_{v}^{*}$}
\State $\mat{Q}_{v}(t+1)=\alpha \bar{\mat{A}} \mat{Q}_{v}(t)+(1-\alpha) \mat{Z}$ \Comment{Vertical propagation}
\EndWhile

\While{$\mat{Q}_{h}(t)$ not converged to $\mat{Q}_{h}^{*}$}
\State $\mat{Q}_{h}(t+1)=\alpha \mat{Q}_{h}(t) \bar{\mat{A}}+(1-\alpha) \mat{Q}_{v}^{*}$ \Comment{Horizontal propagation}
\EndWhile

\State Output $\mat{Q}^{*} = \mat{Q}^{*}_{h}$ as the final converged pairwise constraint matrix
\end{algorithmic}
\end{algorithm}

First, for the vertical constraint propagation, we suppose $\mat{Q}_{v}(0) = \mat{Z}$. Using the horizontal iteration equation, we can obtain,
\begin{align}
\mat{Q}_{v}(t)=(\alpha \bar{\mat{A}})^{t-1} \mat{Z} +(1-\alpha) \sum_{i=0}^{t-1}(\alpha \bar{\mat{A}})^{i} \mat{Z}
\end{align}
\noindent Since the propagation parameter $0 < \alpha < 1$ and the eigenvalues of $\bar{\mat{A}}$ are in $[-1, 1]$, the horizontal propagation has a converged solution as below,
\begin{align}
\lim _{t \rightarrow \infty}(\alpha \bar{\mat{A}})^{t-1} & = 0 \\
\lim _{t \rightarrow \infty} \sum_{i=0}^{t-1}(\alpha \bar{\mat{A}})^{i} & = (I-\alpha \bar{\mat{A}})^{-1} \\
\mat{Q}_{v}^{*} = \lim _{t \rightarrow \infty} \mat{Q}_{v}(t) & = (1-\alpha)(I-\alpha \bar{\mat{A}})^{-1} \mat{Z} \label{eq:vertical_propagation}
\end{align}

\noindent Second, the horizontal constraint propagation can be transformed into a vertical propagation problem by a transpose operation,
\begin{align}
\mat{Q}_{h}\T(t+1) = \alpha \bar{\mat{A}} \mat{Q}_{h}\T(t) + (1-\alpha) {\mat{Q}_{v}^{*}}\T
\end{align}
As shown in Eq.~(\ref{eq:vertical_propagation}), the horizontal propagation converges to 
\begin{align}
{\mat{Q}_{h}^{*}}\T = (1-\alpha)(I-\alpha \bar{\mat{A}})^{-1} {\mat{Q}_{v}^{*}}\T
\end{align}

Therefore, the E2CP constraint propagation algorithm has the following closed-form feasible solution,
\begin{align}
\mat{Q}^{*} = \mat{Q}_{h}^{*} &= (1-\alpha) \mat{Q}_{v}^{*}(I-\alpha {\bar{\mat{A}}}\T )^{-1}  \nonumber \\
&= (1-\alpha)^{2} (I-\alpha \bar{\mat{A}})^{-1} \mat{Z}(I-\alpha \bar{\mat{A}})^{-1}
\end{align}

\section{Detailed FLOP/s analysis}

We provided the total GFLOP/s of different diarization systems in Table~\ref{tab:DER_results}, and discussed the results in Section~\ref{sec:comp_cost}. A detailed version of the FLOP/s analysis is provided in Table~\ref{tab:flops_detail_10min} and Table~\ref{tab:flops_detail_1h}, where we break the FLOP/s numbers into different components.

For the speaker turn detection model and the speaker encoder model, the total number of FLOPs is estimated with \href{https://github.com/tensorflow/profiler}{TensorFlow Profiler} by counting \texttt{total\_float\_ops}. For other components, the total number of FLOPs is estimated with \href{https://flozz.github.io/pypapi/}{PyPAPI} by counting \texttt{PAPI\_FP\_OPS}. The denominator of FLOP/s is based on the length of audio being processed.

\section{Open Source Python Implementation}

We provide a Python-based open source library at \url{https://github.com/wq2012/SpectralCluster}, which covers these implementations:

\begin{enumerate}[leftmargin=*]
    \item Refinement operations on affinity matrix~\cite{wang2017speaker}.
    \item Laplacian matrix~\cite{von2007tutorial}.
    \item K-Means with cosine distance.
    \item Auto-tune~\cite{park2019auto}.
    \item Constained spectral clustering with E2CP~\cite{lu2013exhaustive}.
\end{enumerate}

This library can be installed via \texttt{pip}:
\lstset{backgroundcolor=\color{CodeBackground},basicstyle=\footnotesize\ttfamily}
\begin{lstlisting}[frame=single]
pip3 install spectralcluster
\end{lstlisting}

The ``Turn + E2CP + Auto-tune" configuration that produced the best performance in Table~\ref{tab:DER_results} is provided in \texttt{configs.py}. It can be directly used in the example below:

\lstset{backgroundcolor=\color{CodeBackground},basicstyle=\footnotesize\ttfamily}
\begin{lstlisting}[frame=single]
from spectralcluster import configs

labels = configs.turntodiarize_clusterer.predict(
    embeddings, constraint_matrix)
\end{lstlisting}

However, we also want to clarify that our optimal clustering configuration is based on our specific speaker turn detection and speaker encoder models. The clustering configuration may need to be adjusted when using different models. 

\begin{table*}[t]
  \centering
  \caption{GFLOPS/s of each component for different speaker diarization systems after running for 10min.}
    \begin{tabular}{l|l|c|c|c|c|c|c}
    \toprule
    \multirow{2}{*}{System} & \multirow{2}{*}{Method} & 
    Speaker turn & Speaker & Eigen &
    \multirow{2}{*}{E2CP} & Laplacian & \multirow{2}{*}{Total} \\
      & & 
    detection & encoder &  decomposition & & \& K-Means & \\
    \hline
    \multirow{2}{*}{Dense d-vector}
        & Dense             & 0 & 0.42 & 0.43 & 0 & 0.00 & 0.85   \\
        & Dense + Auto-tune & 0 & 0.42 & 4.34 & 0 & 0.00 & 4.76   \\
    \hline
    \multirow{4}{*}{Turn-to-diarize}
        & Turn                    & 0.58 & 0.42 & 0.00 & 0 & 0.00  & 1.00  \\
        & Turn + E2CP             & 0.58 & 0.42 & 0.00 & 0.00 & 0.00  & 1.00  \\
        & Turn + Auto-tune        & 0.58 & 0.42 & 0.03 & 0 & 0.00  & 1.02 \\
        & Turn + E2CP + Auto-tune & 0.58 & 0.42 & 0.03 & 0.00 & 0.00  & 1.02  \\
    \bottomrule
    \end{tabular}
  \label{tab:flops_detail_10min}
\end{table*}

\begin{table*}[t]
  \centering
  \caption{GFLOPS/s of each component for different speaker diarization systems after running for 1h.}
    \begin{tabular}{l|l|c|c|c|c|c|c}
    \toprule
    \multirow{2}{*}{System} & \multirow{2}{*}{Method} & 
    Speaker turn & Speaker & Eigen &
    \multirow{2}{*}{E2CP} & Laplacian & \multirow{2}{*}{Total} \\
      & & 
    detection & encoder &  decomposition & & \& K-Means & \\
    \hline
    \multirow{2}{*}{Dense d-vector}
        & Dense             & 0 & 0.42 & 36.09 & 0 & 0.04 & 36.54   \\
        & Dense + Auto-tune & 0 & 0.42 & 360.92 & 0 & 0.04 & 361.37   \\
    \hline
    \multirow{4}{*}{Turn-to-diarize}
        & Turn                    & 0.58 & 0.42 & 0.18 & 0 & 0.00  & 1.18  \\
        & Turn + E2CP             & 0.58 & 0.42 & 0.18 & 0.00 & 0.00  & 1.18  \\
        & Turn + Auto-tune        & 0.58 & 0.42 & 1.82 & 0 & 0.00  & 2.81 \\
        & Turn + E2CP + Auto-tune & 0.58 & 0.42 & 1.82 & 0.00 & 0.00  & 2.81  \\
    \bottomrule
    \end{tabular}
  \label{tab:flops_detail_1h}
\end{table*}
\vspace{1in}

\end{appendices}

\end{document}